\documentclass[%
 reprint,
 amsmath,amssymb,
 aps,
]{revtex4-1}
\usepackage[colorlinks=true, allcolors=blue]{hyperref}
\usepackage{graphicx}% Include figure files
\usepackage{dcolumn}% Align table columns on decimal point
\usepackage{bm}% bold math
\usepackage{lipsum}
\newcommand{\degree}{\mbox{$^\circ$}}
\newcommand{\gev}{\mbox{$\rm GeV$}}

\newcommand{\gevc}{\mbox{$\rm GeV/c$}}

\newcommand{\gevcsq}{\mbox{$\rm (GeV/c)^2$}}
\newcommand{\heep}{\mbox{$^1$H(e,e$^\prime$p)}}
\newcommand{\deep}{\mbox{D(e,e$^\prime$p)n}}

\newcommand{\vppi}{\mbox{$\vec{p}_i$}}

\newcommand{\ppm}{\mbox{$p_{m}$}}
\newcommand{\vppm}{\mbox{$\vec{p}_{miss}$}}

\newcommand{\qsq}{\mbox{$Q^2$}}
\newcommand{\Qsq}{\mbox{$Q^2$}}

\newcommand{\sh} {/ \hskip-5pt }

\begin{document}

\preprint{NuPar23}

\title{Double Scattering in Deuteron Electrodisintegration}% Force line breaks with 

\author{Werner U Boeglin and Misak M Sargsian}

\affiliation{Florida International University, Miami, FL 33199}%Lines break automatically or can be forced with \\
%\affiliation{%


%\affiliation{
%\affiliation{
%\affiliation{%


\date{\today}% It is always \today, today,

\begin{abstract}
We demonstrate that at sufficiently high energies where the eikonal regime is established for 
hadronic interactions, the double scattering subprocess can be clearly identified and isolated in quasi-elastic 
deuteron electro-disintegration processes.  Comparing theoretical calculations with 
the recent high precision experimental data we present a ``proof of principle" that these 
processes can be used to study advanced issues related to hadron formation in QCD.
In this case, the double scattering represents as a fermi-scale ``detector" which probes   products of 
high $Q^2$ scattering from the bound nucleon through their rescattering from the spectator nucleon in the deuteron.

%An article usually includes an abstract, a concise summary of the work
%covered at length in the main body of the article. 
%\begin{description}
%\end{description}
\end{abstract}

\maketitle

%\tableofcontents

\section{Introduction}
\label{Sec:I}
One of the important topics in the exploration of the quantum chromodynamic (QCD) structure of hadrons is understanding  their formation from quark-gluon systems, the role of the confinement  and  color field distribution inside hadrons. The main approach in studying these properties is to consider 
hadron production processes involving nuclei where the nuclear medium is used as a
``detector" in kind,  in probing interaction dynamics related to the discussed above phenomena.

One of the best known examples of this strategy is the heavy-ion collision\cite{PHENIX:2004vcz,ATLAS:2008xda} in pursuing the detection of quark-gluon plasma or 
probing hadronization in deep-inelastic  processes in AA, pA (see e.g. \cite{Brodsky:1988xz,Geiger:1992cm,He:2019vgs}) and eA (see e.g. \cite{HERMES:2007plz,Kopeliovich:2003py}) scatterings.
 Studies  of heavy ion  and pA scattering processes have been carried 
out for many decades resulting  in a significant progress in understanding of many aspects of 
QCD dynamics of hadron formation. However the progress was made only in those instances in which the precise knowledge of nuclear structure to interpret the data was not essential.  This is associated with the complication of  accounting for nuclear effects in case of  medium and heavy  nuclei.

In this work we demonstrate that the most simple nucleus, the deuteron, can be used to perform some of 
the explorations of QCD dynamics of baryons. The advantage in this case is  that one can calculate the nuclear effects with sufficiently high accuracy.
The idea that the deuteron can be used to probe QCD dynamics of hard processes by focusing on the rescattering (or double scattering)  subprocess  in high $Q^2$ deuteron electro-disintegration, was suggested in the 1990s~\cite{Frankfurt:1994kt,Frankfurt:1994kk,Laget:1998wu}.
The idea was do use the energy dependence of double scattering strength as a test for the onset of  color transparency~(CT) phenomena. Very recently a dedicated experiment\cite{HollyDoubleCT23} has been approved by the Program Advisory Committee of Jefferson Lab which will probe the onset of CT for double scattering processes for  up to $Q^2=14$~(GeV/c)$^2$. 

However the concept  that  double scattering  can be clearly isolated  and is sensitive to the energy dependence of the amplitude of $pn$-scattering  was never proven experimentally for a benchmark measurement in which one expects the dominance of the hadronic picture in the rescattering (i.e. no CT effects). To prove this concept as a baseline for probing novel QCD effects one needs $Q^2$ large enough 
that  it would be  possible to clearly isolate   the (double scattering) process in which  the produced  nucleon in the $\gamma^*N$ interaction  in the deuteron re-scatters off the spectator nucleon. On the other hand  $Q^2$ needs to be moderate enough that the hadronic picture of rescatteirng is valid and no CT effects are expected.
Once such a subprocess  and its energy sensitivity to $pn$-scattering is established,  
one can extend these studies to higher $Q^2$ to explore the onset of hard  QCD dynamics.

In view of such a program, only recently the first high accuracy measurements at moderately high $Q^2$ became available  that the above mentioned concept could be verified. 

In this work we  use recent progress in both theoretical and experimental studies of  deuteron electrodisintegration processes in $Q^2\sim few$~(GeV/c)$^2$ kinematics to check; (I) if the double scattering 
subprocess can be isolated and described and (II) if the 
energy dependence of $pn$ rescattering, in which the one (fast) nucleon is produced 
by the $\gamma^* N$  interaction  off the bound nucleon and the other nucleon is the spectator in the deuteron, reflects the  energy dependence of the free $pn$-scattering amplitude known independently from $pn\to pn$  elastic scattering experiments.

In Section II we discuss briefly the theory of high  $Q^2$ electrodisintegration of the deuteron. Then in Sec. III
we  focus on kinematics maximally sensitive to the double scattering subprocess and 
  investigate the sensitivity of the electro-disintegration reaction  in this case to  the amplitude of high energy pn scattering.   In Section IV we present the discussion of the 
recent experiments of deuteron electro-disintegration at moderately large $Q^2$ 
performed at Jefferson Lab, focusing again on double scattering kinematics.
 In section V we present the comparison of theory with  experiment in double scattering kinematics 
and  demonstrate that the available data already show the sensitivity of 
the cross section to the energy dependence of the proton-neutron scattering amplitude  thus providing a 
proof of the principle of isolating and probing the  subprocess  of  high energy pn resccattering in the deuteron. 
In Sec.VI we present conclusions and discuss how these 
types of processes can be used in studies of color transparency as well as hadronization processes in the
high $Q^2$ limit.

\section{High {$Q^2$} electro-disintegration of  the deuteron in Generalized Eikonal Approximation}  
\label{Sec:II}

We are discussing the process of  exclusive deuteron electro-disintegration:
\begin{equation}
e+d \rightarrow e^\prime + N_f + N_r,
\label{reaction}
\end{equation}
at large four-momentum transfer, $Q^2 \equiv  {\bf q}^2 - q_0^2$,  kinematics, where ${\bf q}$ and $q_0$ are three-momentum and energy of virtual photon.  
We assign to the struck ~($N_f$)  and recoil~($N_r$) nucleons four-momenta $(E_f,{\bf p_f}$) and ($E_r,{\bf p_r}$) respectively.  For specificity we also consider the struck  nucleon to be a proton and the recoil  a neutron.  One also introduces the missing momentum as; ${\bf p_m} \equiv {\bf q} -  {\bf p_f}$
Note that  for the considered reaction, ${\bf p_m} = {\bf p_r}$.

The  kinematics are such that   one can clearly identify the struck nucleon from the recoil nucleon    with $p_f\sim q \gg p_r\sim m_N$.
The latter, together with the requirement, $Q^2\gg m_N^2$ provides a high energy condition for the process in which  one can treat  $\frac{q_0 -q_z} { q_0+q}\ll 1$ and  $\frac{E_{f}-p_{f,z}}{E_f+ p_{f,z}}\ll 1$ as small parameters.
The emergence of such small parameters allows one to prove the reduction theorem\cite{Sargsian:2001ax}  in which case all possible final state interactions (FSI) of outgoing 
nucleons are ordered in such a way that they can be 
summed up into the single off-shell phenomenological $pn$ scattering amplitude. This makes it possible to 
formulate an effective Feynman diagrammatic rules\cite{Sargsian:2001ax}, using which allows one to perform calculations in a self-consistent way taking into account the relativistic kinematics of the scattering process.  
\begin{figure}[t]
\includegraphics[width=0.48\textwidth]{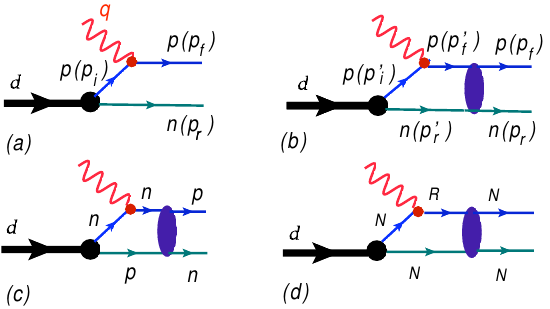}% Here is how to import EPS art
\caption{Diagrams of electrodisintegration amplitude.}
\label{fig:diagrams}
\end{figure}

The above discussed theoretical simplifications created a new opportunity for investigating
reaction(\ref{reaction}) in the high energy limit resulting in intensive theoretical efforts to calculate these processes considering also polarization degrees of freedom (see. e.g. Refs.
\cite{Jeschonnek:2008zg,Ford:2013uza,CiofidegliAtti:2004jg,Laget:2004sm,Sargsian:2009hf}).

In  the discussed high energy limit it can be shown\cite{Sargsian:2009hf,Boeglin:2015cha} that  only four diagrams contribute to the process of Eq.(\ref{reaction}), representing 
plane wave impulse approximation~(PWIA)(Fig.\ref{fig:diagrams}(a)), 
contribution with  final state $pn\to pn$  rescattering~(FSI)  (Fig.\ref{fig:diagrams}(b)), 
contribution with charge interchange $pn\to np$ 
rescattering (ChEx-FSI),(Fig.\ref{fig:diagrams}(c)) and with $RN\to pn$ rescattering (Fig.\ref{fig:diagrams}(d)) in which $R$ represents the resonance produced from electron scattering 
off the bound nucleon.  Other contributions including meson-exchange currents, and spectator 
production in which nucleon $N_f$ is the recoil rather than the struck nucleon are parametrically small. We represent the four dimensional scattering amplitude $A^{\mu}$ through the sum of the above contributions as follows:
\begin{eqnarray}
&&\langle s_f,s_r \mid A^\mu\mid s_d\rangle  =      
\langle s_f,s_r \mid A_0^\mu\mid s_d\rangle + \langle s_f,s_r \mid A_1^\mu\mid s_d\rangle  \nonumber \\
&& \ \   \  \ \ \  + \  \langle s_f,s_r \mid A_{1,chex}^\mu\mid s_d\rangle  
+ \langle s_f,s_r \mid A_{1,R}^\mu\mid s_d\rangle, 
\label{amplitude}
\end{eqnarray}
where $s_d$, $s_r$ and $s_f$ represent the spin states of  the initial deuteron, the final recoil and  the struck 
nucleon, respectively. The amplitudes $A_0$, $A_1$, $A_{1,chex}$ and $A_{1,R}$ correspond to the diagrams of 
PWIA, single diagonal, charge-interchange and $RN\to NN$ rescattering contributions.

With the scattering amplitude defined this way the  unpolorized differential cross section for the 
process~(\ref{reaction}) can be written as follows:
\begin{eqnarray}
&&\frac {d\sigma}{dE_e^\prime, d\Omega_{e^\prime} d p_f d\Omega_f} = \nonumber \\
&&\frac { \alpha^2E^\prime_e}{q^4 E_e} 
\cdot {1\over 6} \sum\limits_{s_f,s_r,s_d,s_1,s_2} \frac{\mid J_e^\mu J_{d,\mu}\mid^2}{2M_d E_f}{p_f^2\over 
\mid {p_f\over E_f} + {p_f - q cos(\theta_{p_f,q})\over E_r}\mid}\nonumber \\
\label{crs}
\end{eqnarray}
where $E_e$ and $E^\prime_e$ are energies of the incoming and scattered electrons, and $M_d$ is the mass of 
the deuteron. Here the leptonic current is defined as $J_e^\mu = \bar u(k_2,s_2)\gamma^\mu u(k_1,s_1)$, with $u(k,s)$ being the Dirac bi-spinors.  The electromagnetic transition current of the deuteron is defined as
\begin{equation}
J_d^\mu = { \langle s_f,s_r \mid A^\mu\mid s_d\rangle\over \sqrt{2(2\pi)^3 2 E_r}},
\label{Jd}
\end{equation}
where $\langle s_f,s_r \mid A^\mu\mid s_d\rangle$ presented in Eq.(\ref{amplitude}).

Further simplification is achieved if one chose the kinematics away from the threshold of resonance, R production. 
In this case only PWIA and FSI and CHEX-FSI contributions define the cross section of the deuteron electrodisintgration.

\subsection{PWIA Contribution}

In the calculation of the PWIA term for not extremely large internal momenta of the deuteron $p_i \le 600$~MeV/c, in which  case $Z$-graph contributions  are small correction the virtual nucleon approximation is justified\cite{Sargsian:2009hf}. 
Within the  effective Feynman diagrammatic approach\cite{Sargsian:2001ax} 
one uses diagrammatic rules to obtain:
\begin{eqnarray}
&& \langle s_f,s_r \mid A_0^\mu\mid s_d\rangle =\nonumber\\ 
&&  -{\bar u(p_r,s_r)}\Gamma_{\gamma^* p}^\mu 
\frac{\sh p_i + m}{ p_i^2-m^2} \cdot \bar u(p_f,s_f)\Gamma_{DNN}\cdot \chi^{s_d},
\label{A0_1}
\end{eqnarray}
where $\Gamma_{\gamma^* p}$ is the electromagnetic vertex of the $\gamma^*N\rightarrow N$ scattering and the vertex 
function  $\Gamma_{DNN}$ describes the transition of the deuteron into the $pn$ system.  Then projecting the deuteron transition vertex to the positive energy solution of the bound nucleon propagator and considering the spectator nucleon as on-shell, one introduces  the 
deuteron wave function\cite{Gribov:1968gs,Bertuchi:1972} in the form:
\begin{equation}
\Psi_d^{s_d}(s_i,p_i,s_r,p_r) = - \frac{ \bar u(p_i,s_i)\bar u(p_r,s_r)\Gamma_{DNN}^{s_d} \chi_{s_d}} 
{(p_i^2 - m^2) \sqrt{2}\sqrt{(2\pi)^3 (p_r^2+m^2)^{1\over 2}}},
\label{wf}
\end{equation}
in which  the bound nucleon is off-energy shell with  $E_i^{off}= M_d-\sqrt{m_N^2+p_r^2}$ and initial momentum ${\bf p_i} = - {\bf p_m}$.
To treat the off-shellness  of the  bound nucleon in the $\gamma^*N$ scattering one 
splits the initial nucleon propagator into on- and off-shell parts that yield on- and off- shell components of  the electromagnetic current for interaction with the incoming electron 
(for details see Ref.\cite{Sargsian:2009hf}).  This results in:
\begin{eqnarray}
& &  \langle s_f,s_r \mid A_{0}^\mu\mid s_d\rangle  =  \sqrt{2}\sqrt{(2\pi)^3 2 E_r}\times \nonumber \\
&& \ \ \ \ \ \ \ \ \ \ \sum\limits_{s_i} J_{N}^\mu(s_f,p_f;s_i,p_i)
\Psi_d^{s_d}(s_i,p_i,s_r,p_r),
\label{A0_3}
\end{eqnarray}
where 
\begin{eqnarray}
& & J_{N}^\mu(s_f,p_f;s_i,p_i)  = \nonumber \\
& & \ \ \ \ \ \ \ \ \ \  J_{N,on}^\mu(s_f,p_f;s_i,p_i)+J_{N,off}^\mu(s_f,p_f;s_i,p_i),
\label{J_N}
\end{eqnarray}
in which on- and off- shell components of electromagnetic current of the bound nucleon are defined as follows:
\begin{eqnarray}
J_{N,on}^\mu(s_f,p_f;s_i,p_i) & = &   \bar u(p_f,s_f)\Gamma^\mu_{\gamma^*N} u(p_i,s_i)\nonumber \\
J_{N,off}^\mu(s_f,p_f;s_i,p_i) & = & \bar u(p_f,s_f)\Gamma^\mu_{\gamma^*N} \gamma^0 u(p_i,s_i)\times\nonumber \\
& & \frac {E_i^{off} - E_i^{on}}{2m},
\label{J_on_off}
\end{eqnarray}
with the electromagnetic vertex expressed through two independent Dirac and Pauli form-factors:
\begin{equation}
\Gamma^\mu = F_1(Q^2)\gamma^\mu + {F_2(Q^2) \over 2m}i \sigma^{\mu,\nu}q_\nu.
\label{Gamma_N}
\end{equation}

\subsection{FSI Contribution}
The FSI and ChEx-FSI diagrams in Fig.\ref{fig:diagrams}(b) and (c) are calculated within the Generalized Eikonal Approximation\cite{Frankfurt:1996xx,Sargsian:2001ax,Sargsian:2009hf} in which using 
effective Feynman diagrammatic rules\cite{Sargsian:2001ax} one calculates the scattering amplitude as:
\begin{eqnarray}
& & \langle s_f,s_r \mid A_{1}^\mu\mid s_d\rangle  = -\int {d^4p_r^\prime \over i (2\pi)^4}  \nonumber \\
&& \times 
\frac { \bar u(p_f,s_f)\bar u(p_r,s_r) F_{NN}[\sh p^\prime_r + m][\sh p_d - \sh p^\prime_r+\sh q+m]}
{(p_d-p^\prime_r+q)^2-m^2+i\epsilon}
\nonumber \\
 &&\times \frac {\Gamma_{\gamma^*N}[\sh p_d-\sh p^\prime_r + m]\Gamma_{DNN}\chi^{s_d}}{((p_d-p^\prime_r)^2-m^2+i\epsilon)
(p^{\prime 2}_r - m^2 + i\epsilon)},
\label{A1_0}
\end{eqnarray}
were $F_{NN}$ represents the invariant off-shell direct $pn\rightarrow pn$  or charge-interchange 
$pn\rightarrow np$ scattering amplitude, that can be expressed as follows:
\begin{equation}
F_{NN}(s,t) = \sqrt{s(s-4m^2)}f_{NN} (s,t), 
\end{equation}
with $s = (q+p_d)^2 = (p_f + p_r)^2$ and $t=(p^\prime_r -p_r)^2$. 
Here $f_{NN}$ represents the parametrization of the  $pn$ direct and charge-exchange scattering amplitudes in  the diffractive form such that  for direct scattering $Im f_{NN}=\sigma_{pn}^{tot}$, where 
$\sigma_{pn}^{tot}$ is the total $pn$ scattering cross section.

 Furthermore, in Eq.(\ref{A1_0}) we first, integrate by $d^0p_{r^\prime}$ through the positive energy pole of 
 the spectator nucleon in the intermediate state, 
 then integrate by $p^\prime_{r,z}$  splitting the propagator of the struck nucleon in the intermediate 
 state into  pole and principal-value parts.  This results in:

\begin{widetext}
\begin{eqnarray}
& & \langle s_f,s_r \mid A_{1}^\mu\mid s_d\rangle  
  =     {i \sqrt{2}(2\pi)^{3\over 2}\over 4} \sum\limits_{s^\prime_f,s^\prime_r,s_i} 
\int {d^2p_r^\prime \over  (2\pi)^2}  \frac{\sqrt{2\tilde E^\prime_r}\sqrt{s(s-4m^2)}} {2\tilde E^\prime_r |q|}  
%\times \nonumber \\
%
\langle p_f,s_f;p_r,s_r\mid f^{NN,on}(t,s)\mid \tilde p^\prime_r,s^\prime_r; \tilde p^\prime_f,s^\prime_f\rangle \nonumber \\
&& \ \ \ \ \ \ \ \ \ \  \times  J_{N}^\mu(s^\prime_f,p^\prime_f;s_i,\tilde p^\prime_i) \cdot  \Psi_d^{s_d}(s_i,\tilde p^\prime_i,s^\prime_r,\tilde p^\prime_r) 
\nonumber  \\
& & \ \ \ \ \ -  {\sqrt{2}(2\pi)^{3\over 2}\over 2} \sum\limits_{s^\prime_f,s^\prime_r,s_i} {\cal P}\int {dp^\prime_{r,z}\over 2\pi} 
\int {d^2p_r^\prime \over  (2\pi)^2}  \frac{\sqrt{2E^\prime_r}\sqrt{s(s-4m^2)}} {2E^\prime_r |{\bf q}|} 
%\times \nonumber \\
{\langle p_f,s_f;p_r,s_r\mid f^{NN,off}(t,s)\mid p^\prime_r,s^\prime_r;p^\prime_f,s^\prime_f\rangle
\over  p^\prime_{r,z}- \tilde p^\prime_{r,z} } \nonumber \\
& & \ \ \ \ \ \ \ \ \ \ \times J_{N}^\mu(s^\prime_f,p^\prime_f;s_i,p^\prime_i)
\cdot  \Psi_d^{s_d}(s_i,p^\prime_i,s^\prime_r,p^\prime_r),
\label{a1_fsi}
\end{eqnarray}
\end{widetext}
where $\tilde p^\prime_r = (p_{r,z}-\Delta, p^\prime_{r,\perp})$, $\tilde E^\prime_r = \sqrt{m^2 + \tilde p^{\prime 2}_r}$, 
$\tilde p^\prime_i = p_d - \tilde p^\prime_r$ and $\tilde p^\prime_f = \tilde p^\prime_i + q$ and 
\begin{equation}
\Delta = {q_0\over |{\bf q}|}(E_r - E^\prime_r) + {M_d\over |{\bf q}|}(E_r - E^\prime_r)  + {p^{\prime 2}_r-m^2\over 2  |{\bf q}|}.
\label{Delta}
\end{equation}
Note that while $f^{NN,on}(t,s)$ is taken from experiments the off-shell counterpart is modeled according to the relation:
\begin{equation}
f^{NN,off} = f^{NN,on} e^{B(m_{off}^2 -m^2)},
\label{fnn_off}
\end{equation}
where $m_{off}^2 \equiv (p^\prime_f)^2$.
As it follows from Eq.(\ref{a1_fsi}) the principal value part of FSI contributes to the real part of the rescattering amplitude which is small in high energy limit. As a result the accuracy of the approximation in Eq.(\ref{fnn_off}) is not expected to render large uncertainities in 
the overall estimate of the rescattering contribution.

Regarding the $f^{NN,on}$ amplitude one obtains it from the analysis of experiments in direct $pn\rightarrow pn$ scattering for 
evaluating FSI and $pn\rightarrow np$ scattering for evaluating ChEx-FSI.  It is worth mentioning that  
$pn\rightarrow np$ scattering is dominated by pion exchange. As a result its amplitude decreases with increasing energy by a factor of $\sqrt{s}$ compared to the direct $pn\rightarrow pn$ scattering. The latter, at considered energies, in forward direction is  dominated by Pomeron exchange with a weak energy dependence  and is predominantly imaginary.   Thus, the pole term in Eq.(\ref{a1_fsi}) has a dominating contribution from the direct  $pn\rightarrow pn$ scattering amplitude

\section{Double Scattering Contribution}
\label{Sec:III}

\begin{figure}[htb]
\includegraphics[width=0.40\textwidth]{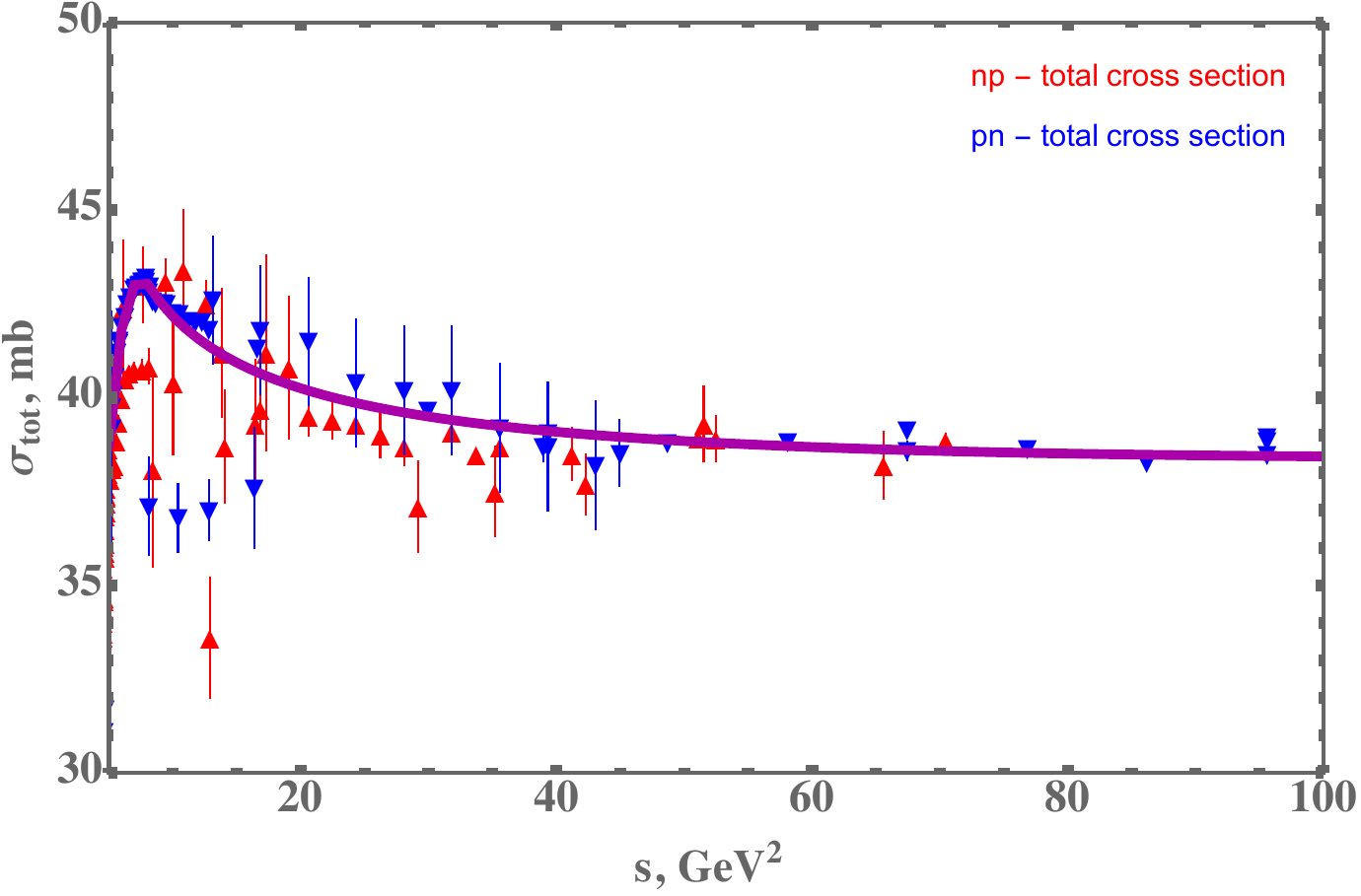}% 
\caption{ Comparison of the parametrization of $\sigma_{pn,tot}$  with the experimental data \cite{ParticleDataGroup:2022pth} as a function of $s$.}
\label{fig:sigma_pn_tot}
\end{figure}

The total invariant energy of the reaction (\ref{reaction}) is defined as
\begin{equation}
s \equiv (q+p_d)^2 = M_d^2 + Q^2({M_d\over m_N x}-1),
\label{s}
\end{equation}
which indicates that it linearly grows with $Q^2$.  For sufficiently large $s$ 
one expects that the direct FSI amplitude defines the whole dynamics of the final state interaction 
in the reaction (\ref{reaction}). In this case the FSI amplitude can be parameterized as:
\begin{equation}
    f_{pn}(s,t) = (\sigma_{pn,tot}(s)(i + \alpha(s))e^{-{B(s)\over 2}t},
\label{fnn}
\end{equation}
where the real part of the amplitude $\alpha(s)$ is a small correction in the high energy limit. The total $pn$ scattering 
cross section is defined  by $\sigma_{pn,tot}(s)$. These and other parameters $\alpha(s)$ and $B(s)$ are defined from 
the $pn\to pn$ scattering experiments.   The comparison of the fit of $\sigma_{pn,tot}$ with 
available data\cite{ParticleDataGroup:2022pth} is shown in Fig.\ref{fig:sigma_pn_tot} which indicates noticeable dependence on 
the invariant energy  up to $s\le 20$~GeV$^2$, the range within which current experiments of reaction (\ref{reaction}) are
performed.

\begin{figure}[htb]
\includegraphics[width=0.48\textwidth]{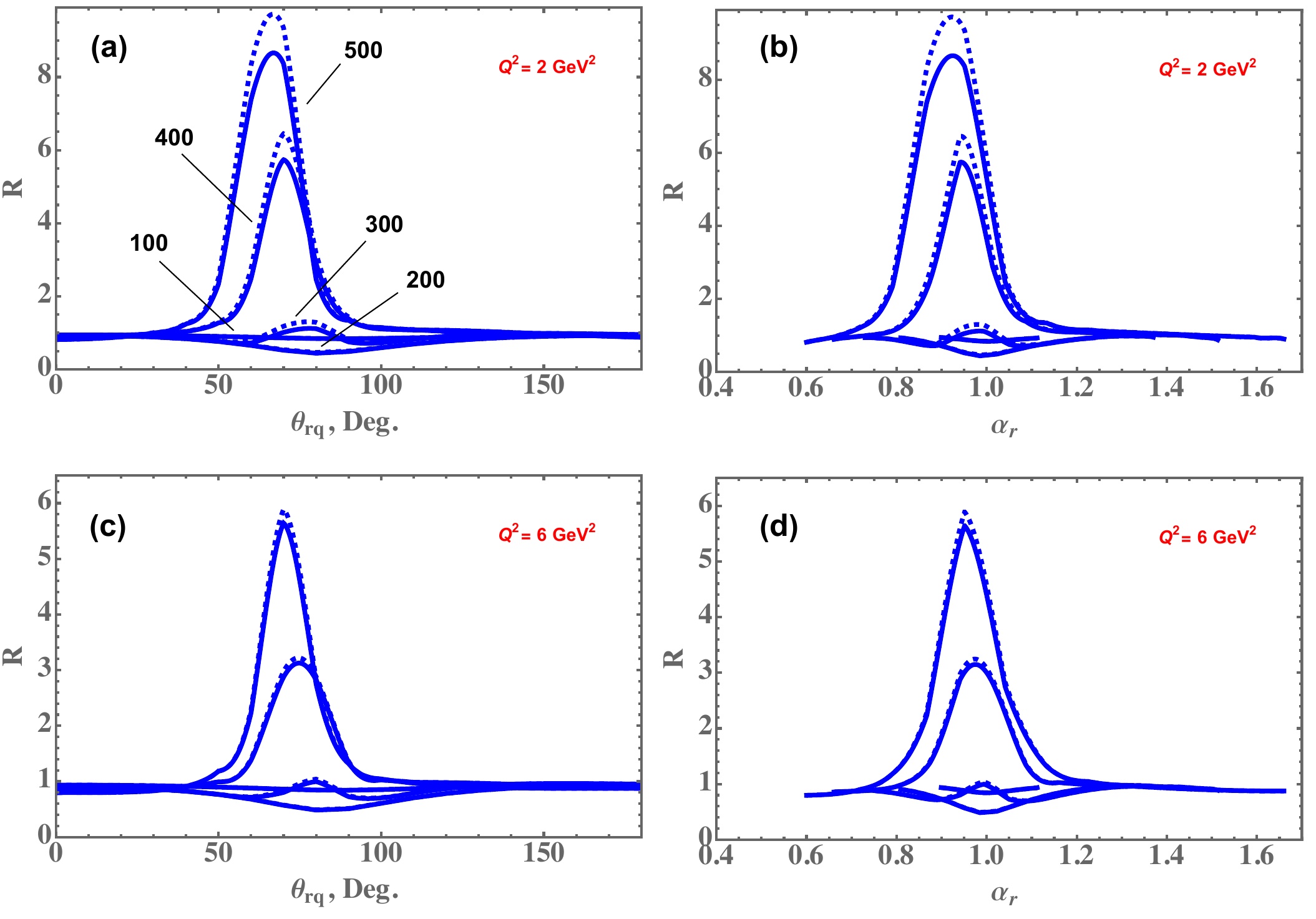}% 
\caption{ The dependence of $R$ on the angle of recoil nucleon for different values of recoil momenta (a) and (c) or different values of light-front momentum fraction of recoil nucleon (b) and (d). Solid curves corresponds to calculations that include PWIA and FSI terms, while dashed line include also ChEx-FSI.}
\label{fig:Rprediction}
\end{figure}

Because of the predominantly imaginary nature of the $f_{pn}$ amplitude the cross section of the reaction (\ref{reaction}) is proportional to the following combination of the scattering amplitudes:
\begin{equation}
\mid A\mid ^2  = \mid A_{PWIA}\mid^2 - 2{\cal R}e A_{PWIA}A_{FSI} + \mid A_{FSI}\mid^2.
\label{A2}
\end{equation}
Here with the increase of missing momentum in the reaction $A_{PWIA}$ decreases much faster than $A_{FSI}$, since for 
the former the wave function is defined by the magnitude of the recoil momentum $p_r$ while for the latter it is an integral 
in Eq.(\ref{a1_fsi}) which  is dominated at smaller values of momenta entering the wave function of the deuteron. 
As a result if we construct a ratio:
\begin{equation}
R_{PWIA} = {\sigma_{Full}\over \sigma_{PWIA}},
\label{R}
\end{equation}
where $\sigma_{PWIA}$ is the cross section (\ref{crs}) calculated by PWIA amplitude only and 
$\sigma_{Full}$ includes also rescattering amplitudes, then  
at small $p_r$ the ratio is close to unity; then it decreases with an increase of $p_r$ due to the increasing contribution of the interference 
term in Eq.(\ref{A2}).  However, with the further increase of $p_r$,  the dominating term becomes the 
term proportional to  $\mid A_{FSI}\mid^2$ (hereafter referred to {\em double scattering term}). This 
dominance reflects the fact that the interference term is proportional to the deuteron wave function with the argument of $p_r$ while the double scattering term is defined only  by the integral~(\ref{a1_fsi}), in which deuteron momentum is integrated over the transverse component of the momentum transfer in the rescattering. As a result, the further increase of  $p_r$ 
will result in the diminishing interference term, while the   $\mid A_{FSI}\mid^2$ changes slowly.
Thus one expects strong dependence of the ratio $R$ as a function of $p_r$ in the kinematics 
dominated by final state reinteractions.
This pattern for $R$ is predicted in Refs.\cite{Frankfurt:1996xx,Sargsian:2001ax,Sargsian:2009hf} presented in Fig.\ref{fig:Rprediction}.
As one can see in the figure the FSI is large  at transverse angles $\theta_{rq}\sim  70^{\degree}-80^{\degree}$ with the double scattering term  dominating at  $p_r\gtrsim 400$~MeV/c at $\theta_r\approx 70^0$.  The latter corresponds to the situation in which the momentum entering in the  wave function in the integrand  of the pole part of  
Eq.(\ref{a1_fsi}) is very small, maximizing the FSI amplitude. 
 In many cases, it is convenient to consider the FSI contribution as a function of the light cone momentum fraction of the deuteron carried by the recoil nucleon:  $\alpha_r  = {E_r-p_{r,z}\over m_N}$.   The choice of $\alpha_r$ as a kinematic parameter is associated with the fact that it provides a Lorentz boost invariant condition for maximal FSI effect at $\alpha_r\approx 0.9-1$ (see Fig.\ref{fig:Rprediction}(b)(d)).
 
If the cross section of reaction (\ref{reaction}) at above discussed kinematics  ($p_r\ge 400$~MeV/c and $\alpha_r\approx 0.9-1$)   is defined predominantly by the double scattering term and the all above mentioned effects, like principal value and charge-interchange FSI contributions are small at high energy, then one expects 
that the $s$-dependence of the ratio $R$~(\ref{R}) should reflect the $s$ dependence of $\mid f_{pn}(s,t)\mid ^2$ defined in Eq.(\ref{fnn}).
\begin{figure}[htb]
\includegraphics[width=0.40\textwidth]{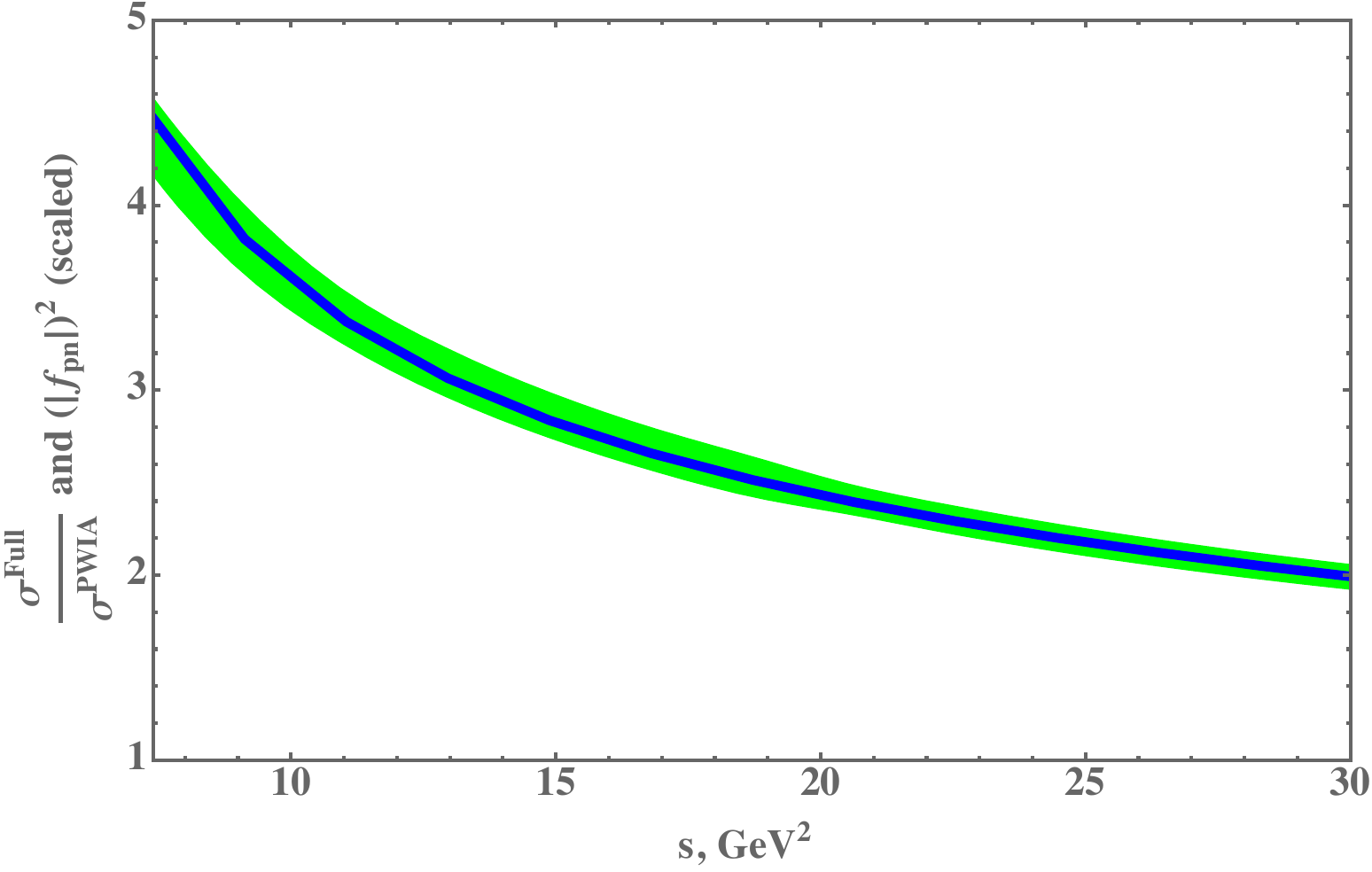}% 
\caption{ Comparison of double scattering contribution to the $\mid f_{pn}(s,t)\mid^2$.}
\label{fig:dsc_fpn2}
\end{figure}
To verify this, in Fig.\ref{fig:dsc_fpn2} we compare the $s$ dependencies of ratio-$R$ and $\mid f_{pn}(s,t)\mid ^2$ (scaled at $s=30$~GeV$^2$) for $\alpha_r=1$ and $p_r=400$~MeV/c  and $\phi_{rq}=\pi$. As the figure shows the double scattering term reproduces the shape of the energy dependence of the modulus square of the $pn$  scattering amplitude very well. 
This comparison shows that the cross section of reaction (\ref{reaction})  at the kinematics of  double scattering can be  used as a tool for investigation of the energy 
dependence of the amplitude of $pn$ elastic scattering in high energy limit. 
This is the main argument for the reaction (\ref{reaction}) to be used for investigation of  the hadronic interactions in the deuteron at much larger $Q^2$. 
%One such example is the   probing color transparency phenomena in high $Q^2$ limit (see e.g. Ref.\cite{Li:2022uvf}).

\section{Experimental status of high $Q^2$ electro-disintegration processes in double scattering kinematics}

Early experiments of reaction (\ref{reaction}) 
carried out with electron accelerators with energies below 1~$\gev$ 
were able to measure coincidence cross sections at large missing momenta, however, these measurements were dominated by 
long range  processes  
that prevented one to 
probe the short range structure of the deuteron 
(for a discussion and  summary of low energy experiments 
see Ref.~\cite{Boeglin:2015cha} and references therein). 
In these experiments
due to the relatively low incident energy and the associated low momentum transfer it was not possible 
for example, 
to 
separate an event where a large initial momentum proton was hit and the spectator neutron recoiled with the equal and opposite momentum from an event where a lower momentum neutron was hit and the recoiling proton was measured. The second contribution can become important when the final state proton momentum is of the same order or even smaller than the missing momentum being probed.
This situation
changed with the operation of Jefferson Lab which allowed experiments using the 6 as well as the 11~GeV 
%CW
high intensity  continuous wave

electron beams. In these experiments final state proton momenta were typically more than 1.5 $\gevc$ which is about a factor three larger than the largest missing momenta probed.

\begin{figure}[htb]
\includegraphics[width=0.40\textwidth]{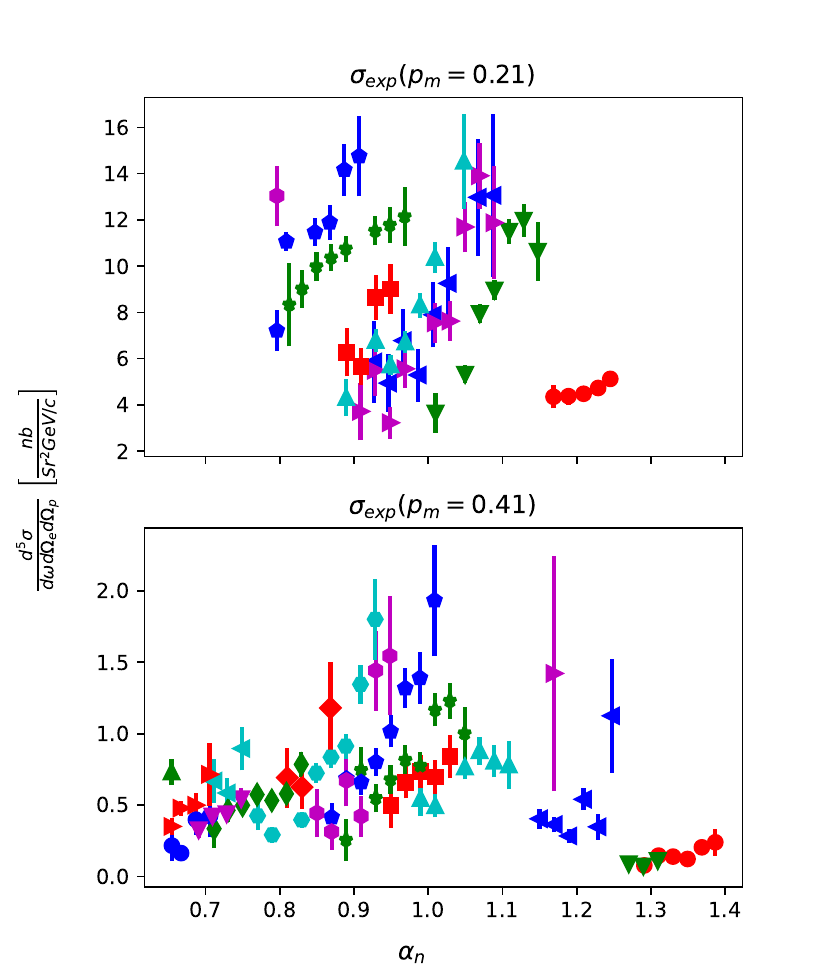}% 
\caption{Experimental cross sections as a function of $\alpha_n$ for $\qsq = 2.1$ $\gevcsq$. Top panel $\ppm = 0.21$ $\gevc$, bottom panel $\ppm = 0.41$ $\gevc$. The different color and symbol combinations correspond to the different contributing spectrometer settings.}
\label{fig:sig_exp_21_41}
\end{figure}
Experiments carried out in Hall~A~\cite{Boeglin:prl11} and at CLAS in Hall~B~\cite{Egiyan:2007qj}  
of Jefferson Lab, measuring 
$\deep$ cross sections for missing momenta up to 0.5~$\gevc$ over a range of neutron recoil angles from $\sim 10\degree$ up to almost $180\degree$ with 
momentum transfers being order of {\em several}~$\gevcsq$.
Especially the Hall~A experiment with its fine kinematic binning 
allowed
one to perform a detailed study of the $\deep$ reaction for different, well-defined kinematic settings. Missing momentum distributions for the high momentum transfer data-set ($\Qsq = 3.5$ $\gevcsq$) have been published\cite{Boeglin:prl11} together with recoil angular distributions of the neutron for missing momenta of $\ppm = 0.2, 0.4$ and $0.5$~$\gevc$.  Both,
Hall A and Hall B 
experiments demonstrated that at 
$Q^2\gtrsim 2$~$\gevcsq$ FSI are strongly 
anisotropic in the angle of the recoil neutron ($\theta_{nq} $)
 with respect the 3-momentum transfer, $\bf q$ peaking at 
  $\theta_{nq} \sim  70{\degree}$ and being small 
 at parallel and antiparallel directions.
 It is worth mentioning that at lower $Q^2$ ($\qsq = 0.8~\gevcsq$) 
 large FSI contributions were found for all recoil angles for missing momenta of 0.4 $\gevc$ and above~\cite{HKhanal2014}. 

 As it was discussed in Sec.~\ref{Sec:III},
 the above mentioned anisotropy of FSI is  the signature of 
 the onset of the high energy regime and was 
 predicted by eikonal  based models such as the generalized eikonal approximation~\cite{Frankfurt:1994kt,Frankfurt:1996xx,Sargsian:2001ax} 
 as well as models based on similar high energy approximations\cite{Jeschonnek:2008zg,Ford:2013uza,CiofidegliAtti:2004jg,Laget:2004sm,Sargsian:2009hf}. 
 
 This 
 anisotropy
 made it possible to select kinematic regions where FSI are reduced, providing a more direct access to the underlying momentum distribution in 
 the deuteron.
 Probing a ``genuine" momentum distribution in the deuteron by selecting 
 kinematics of reduced FSI was the goal of the first 
 measurement\cite{Boeglin:prl11} as well as a recent measurement
 at  larger $\qsq = 4.25\gevcsq$ and at
 significantly larger
 missing momenta, up to $1\ \gevc$. 
 It is worth mentioning that 
 the latter experiment, performed 
 in Hall C at Jefferson Lab~\cite{Yero2020}, produced 
 data  that cannot be described by current  theoretical models 
 of the $\deep$ reaction and may indicate an onset of 
 non-nucleonic components in the deuteron at missing momenta above 
 750~MeV/c\cite{Sargsian:2022rmq}.

We now shift our focus to the kinematics dominated by FSI.
As it was  discussed in the introduction (\ref{Sec:I}), our goal is to verify 
whether
isolating the FSI will offer a possibility to  
 probe the nucleon-nucleon scattering amplitude.

\begin{figure}[htb]
\includegraphics[width=0.40\textwidth]{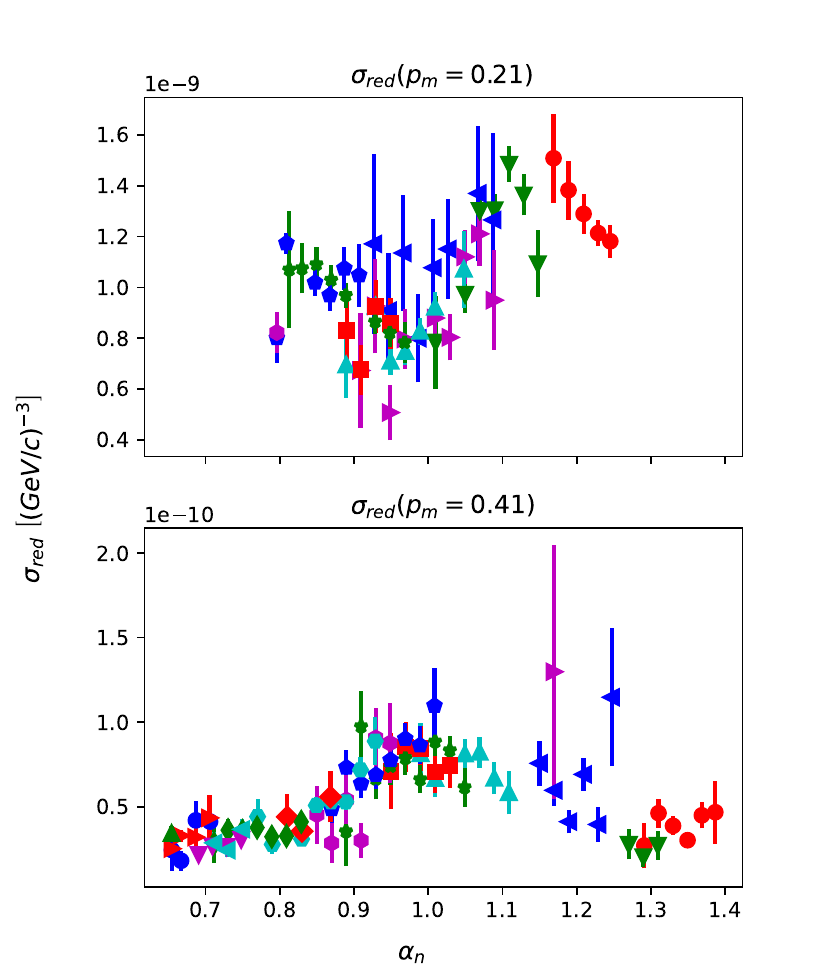}% 
\caption{Experimental reduced cross sections as a function of $\alpha_n$. Top panel $\ppm = 0.21$ $\gevc$, bottom panel $\ppm = 0.41$ $\gevc$. As in Fig.~\ref{fig:sig_exp_21_41}, the different color and symbol combinations correspond to the different contributing spectrometer settings.}
\label{fig:sig_red_21_41}
\end{figure}

For this, according to Sec. \ref{Sec:III} one needs  to 
isolate
the double-scattering contribution to the reaction (\ref{reaction}).
As a first step in this direction we re-analyzed the Hall A data at $\qsq$-values of 2.1 and 3.5~$\gevcsq$ for missing momenta $\ppm = 0.2, 0.4$ and 0.5 $\gevc$
with recoil neutrons produced centered around the $\phi_{nq} = 0^0$ plane. Since we found that the $\ppm = 0.5$ $\gevc$ data are more sensitive to the deuteron wave function, we focus this re-scattering analysis on the missing momentum settings $\ppm = 0.2$ and 0.4 $\gevc$.
The data at $\qsq = 2.1 \pm 0.25 \gevcsq$ (here $\pm 0.25$ indicates the $\qsq$ bin width) were taken at an average beam energy of 4.703~$\gev$ and the ones at $\qsq = 3.5 \pm 0.25 \gevcsq$ were taken at an average beam energy of 5.008~$\gev$. 
The analysis of the two $\qsq$ data-sets, proceeded almost identical to the analysis described in Ref.~\cite{Boeglin:prl11}.  The coincidence efficiencies were determined from $\heep$ measurements and found to be 
$97\pm 5.5 \%$ and $96 \pm 2\%$, for the $\qsq = 2.1$ and $\qsq = 3.5$~$\gevcsq$ data-sets, respectively. For every kinematic bin the systematic error due to uncertainties in the measured kinematic variables was estimated and added in quadrature to the statistical error. An additional estimated global systematic error of 4.0\% was added in quadrature to take into account uncertainties in beam charge measurements, detector efficiencies, target thickness determination, target boiling corrections and the nucleon form factor data. 

\begin{figure}[htb]
\includegraphics[width=0.40\textwidth]{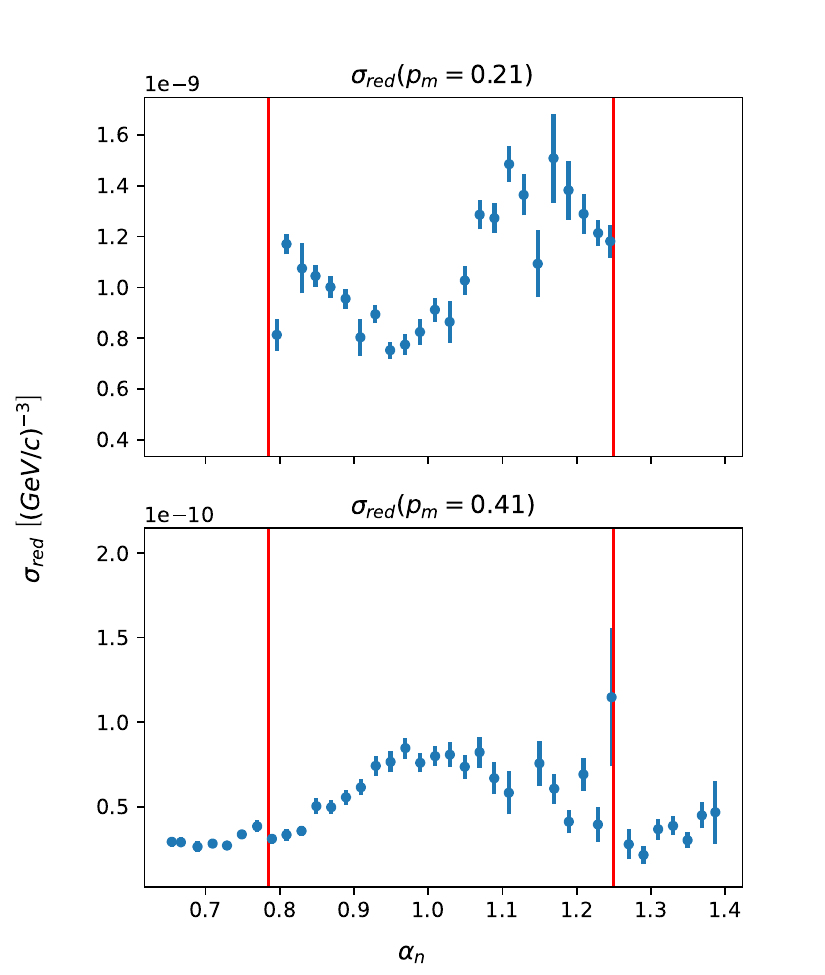}% 
\caption{Average experimental reduced cross sections as a function of $\alpha_n$. Top panel $\ppm = 0.21$ $\gevc$, bottom panel $\ppm = 0.41$ $\gevc$. The red vertical lines indicate the region of overlapping $\alpha_n$ values which will be used in the analysis described below. }
\label{fig:sig_red_avg_21_41}
\end{figure}

We determined absolute cross sections as a function of missing momentum and neutron light-front momentum fraction:
\begin{equation}
\alpha_n = \frac{E_n - p_{n,z}}{M_d/2}
\label{alphan}
\end{equation}
using a missing momentum bin size of $\Delta\ppm = \pm 0.01$~$\gevc$ and a momentum fraction bin size of $\Delta\alpha_n = \pm 0.01$.  
In the above equation $p_{n,z}$ is the component of the recoil neutron momentum in the $\bf q$ direction.
The rationale  for choosing the neutron's light cone momentum fraction as a kinematic parameter was discussed in Sec.\ref{Sec:III}.

As an example,
Fig.~\ref{fig:sig_exp_21_41} shows the measured cross sections for missing momentum values of $\ppm = 0.21$ and 
0.41~$\gevc$. The data are shown in groups of different color and symbol combinations. Data points with the same color and symbol belong to the same spectrometer setting. The measured cross sections for a given value of $\alpha_n$  from different spectrometer settings differ considerably which is due to slightly different kinematic settings (electron scattering angle, proton direction etc.) that result in the same $\alpha_n$ and $\ppm$ bin and it is quite difficult to observe an emerging structure from these cross sections. This situation changes when one 
evaluates the reduced cross sections 
defined  as follows:
\begin{equation}
\sigma_{red} = \frac{\sigma_{exp}}{k \sigma_{ep} f_{rec}}
\label{sig_red}
\end{equation}
where $k$ is a kinematic factor, $\sigma_{ep}$  the calculated
off-shell e-p cross section and $f_{rec}$ a recoil factor. 
The factors are chosen such that 
$\sigma_{red}$ would correspond to the momentum distribution if there were no FSI and no other processes such as meson exchange currents and isobar configuration contributing to the cross section that destroy the relation $\vppi = -\vppm$, where $\vppi$ is the initial momentum of the bound nucleon.

The resulting reduced cross sections as shown in Fig.~\ref{fig:sig_red_21_41} 
significantly diminish
many of the cross section variations in a given $\alpha_n - \ppm$ bin. 
The data-points of the different spectrometer settings are now statistically overlapping and a  
clearer pictures emerges.

Within PWIA,
for constant missing momentum the observed reduced cross section should be independent of $\alpha_n$
and observed deviations at considered $Q^2$ are due to (non-PWIA) processes of 
Fig.(\ref{fig:diagrams}) (b),(c) and (d).
As Fig.\ref{fig:sig_red_21_41} shows  
the observed reduced cross sections 
are not independent of $\alpha_n$. 
Moreover they show pattern similar to Fig.\ref{fig:Rprediction};
a reduction abound 
$\alpha_n \approx 0.95-1.0$ 
for $\ppm = 0.21$ $\gevc$ and a sizable enhancement  at the same $\alpha_n$ value for $\ppm = 0.41$ $\gevc$. 
As a  next step we calculated the weighed mean of all spectrometer contributions for each $\alpha_n$-bin and for each $\ppm$ value. The result is shown in Fig.~\ref{fig:sig_red_avg_21_41}. The vertical red lines indicate the range of common $\alpha_n$ values for the two missing momentum settings allowing for a comparative study of double-scattering in the $\deep$ reaction discussed in the following section. The variations of the reduced cross sections for the two different missing momentum settings are now very clear, 
allowing to identify the domain of maximal FSI.

\section{Comparison of Theory with Experiment}
\begin{figure}[htb]
\includegraphics[width=0.40\textwidth]{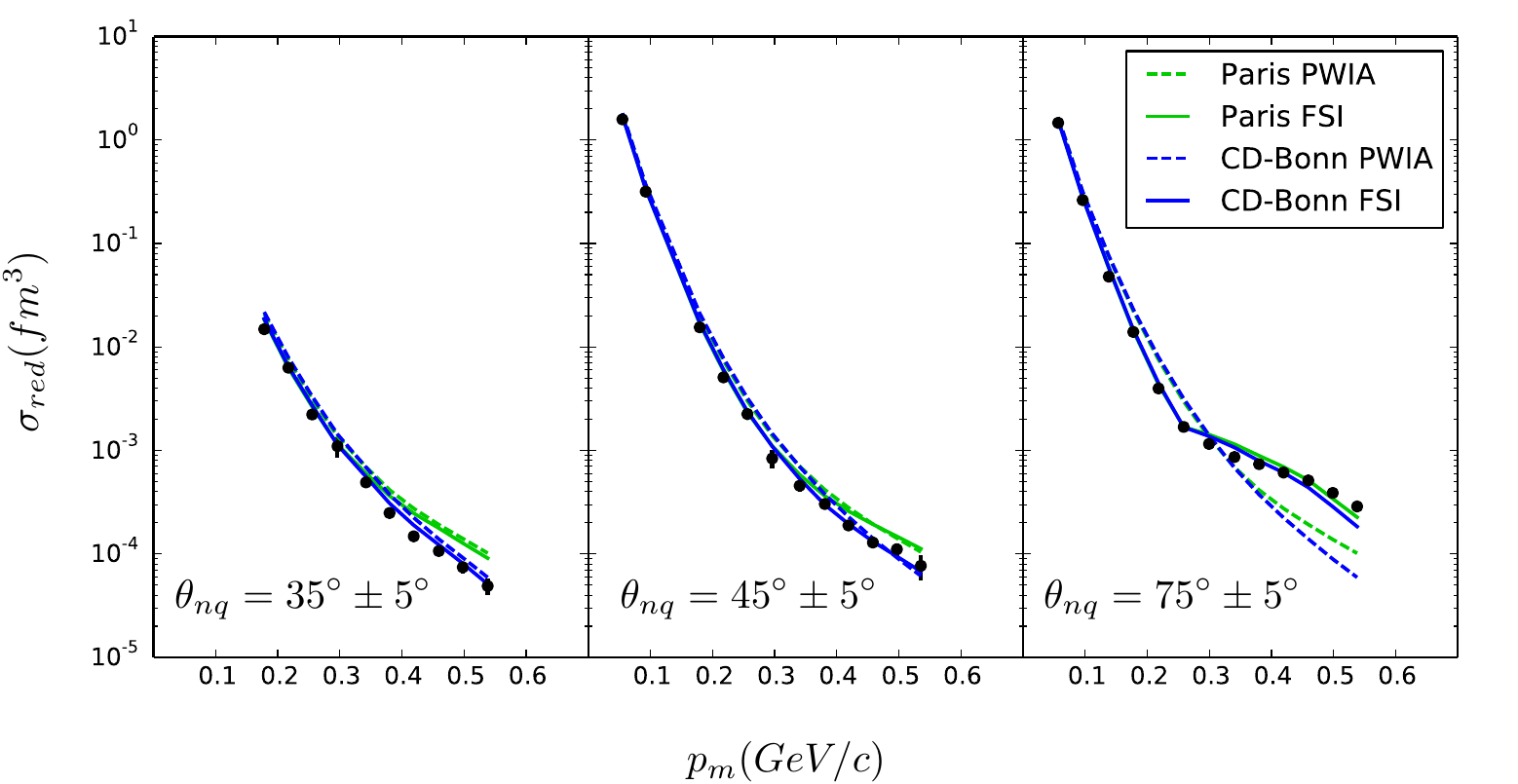}% 
\caption{ Comparision of GEA calculation with reduced experimental cross sections }
\label{fig:pm_distribution}
\end{figure}

As above experimental analysis shows that data clearly indicate the enhanced FSI contribution at $\alpha_n=\alpha_r\approx 1$ and demonstrate 
the pattern predicted in Fig.\ref{fig:Rprediction}. 
In Ref.\cite{Boeglin:prl11}  the first  theoretical comparison was focused on isolating kinematics most sensitive to the genuine deuteron structure,  by minimizing FSI contribution.  As Fig.\ref{fig:pm_distribution} shows, indeed such  kinematics were found at $\theta_{nq}\le 40^0$ in which 
the large sensitivity to the deuteron wave function was observed.
The same comparison (right panel of Fig.\ref{fig:pm_distribution}) also demonstrates the enhanced role of final state interactions at $\theta_{nq}\approx 75^0$. As the latter shows starting at $p_r=p_m\ge 400$~MeV/c the cross section is almost all due to rescattering contribution as it was discussed in Sec.\ref{Sec:III}.

\begin{figure}[htb]
\includegraphics[width=0.40\textwidth]{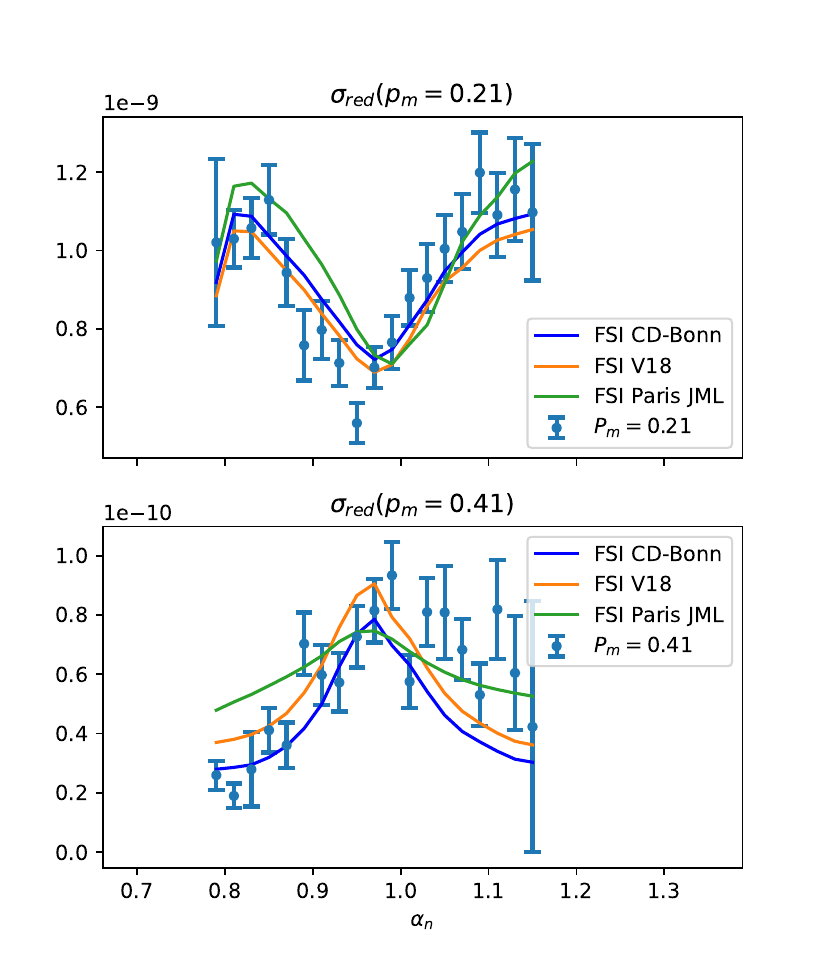}% 
\caption{ Comparison of theoretical calculations with reduced experimental cross sections at
$p_m = 201$~MeV/c and $401$~MeV/c for the range of $\alpha_n$.}
\label{fig:alpha_dependence}
\end{figure}

We focus now on kinematics dominated by final state interaction. In Fig.\ref{fig:alpha_dependence} one observes that  the data unambiguously identify the kinematics dominated by the interference (at $p_R = p_m\approx 200$~MeV/c) and double scattering  (at $p_r= p_m\approx 400$~MeV/c) terms in Eq.(\ref{A2}) (compare with Fig.\ref{fig:Rprediction}). The data also confirms that the  maximum of FSI takes place at $\alpha_n\approx 1$.   

It is worth noting that the underestimation of calculations at 
$\alpha_n> 1.1$ is most probably due to intermediate $\Delta$-isobar production since it corresponds to the kinematics close 
to the threshold of $\Delta$ production in the intermediate state.  The calculations presented in Fig.\ref{fig:alpha_dependence} do not include intermediate $\Delta$ -isobar contribution (similar to Fig.\ref{fig:diagrams}~(d)).

As the lower panel of Fig.\ref{fig:alpha_dependence} shows it has a clear signature of maximal double scattering contribution, at $p_m = 400$~MeV/c, similar to  Fig.\ref{fig:Rprediction} (b) and (d). Thus we attempt to check the validity of the statement that the double scattering term is sensitive to the shape of $\mid f_{pn}\mid^2$ as it was
presented in Fig.\ref{fig:dsc_fpn2}.  
For this we evaluated the experimental ratio R of Eq.(\ref{R}) by dividing the experimental cross sections by PWIA calculations at $p_r=p_m = 400$~MeV/C averaged over $\alpha_n= 1\pm 0.1$, measured at $Q^2=2$ and $3.5$~(GeV/c)$^2$ corresponding to the  total invariant energy of $s=5.8$ and $7.9$~GeV$^2$ respectively. 
In theoretical calculations we choose  the CD-Bonn wave function of 
the deuteron which gave an overall best description of the data for all kinematics of the experiments.  
 \begin{figure}[thb]
\includegraphics[width=0.40\textwidth]{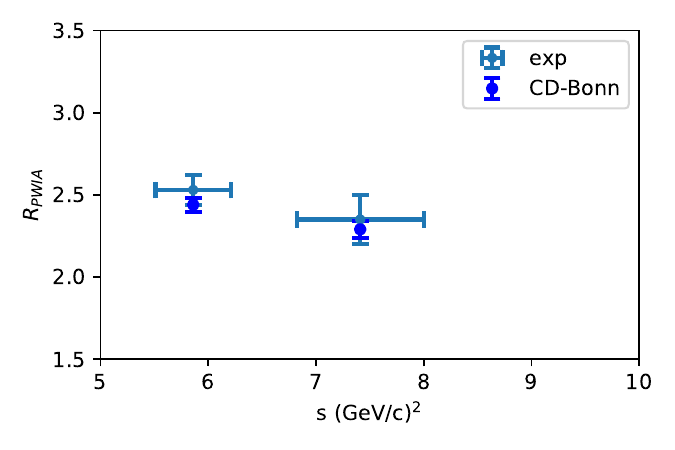}% 
\caption{ Comparison of theoretical calculations with reduced experimental cross section ratios at
$p_m = 410$~MeV/c for the range of $\alpha_n=1\pm 0.1.$.}
\label{fig:sdependence}
\end{figure}

% WB this is not quite clear and needs to be clarfied.
 As Fig.\ref{fig:sdependence} shows despite $s$ being relatively small that ChEx-FSI neglected in Fig.\ref{fig:dsc_fpn2}  is not negligible, the first such comparison indicates that the data reproduce the shape of the $\mid f_{pn}\mid^2$ amplitude that enters in FSI. 
 Note that the ratio in Fig.\ref{fig:sdependence}  for both the 
 experiment and theory  has been determined by fitting a parabola to the peak regions ($\alpha_n = 1\pm 0.1$) and the respective maxima have been determined from the fitting parameters resulting in error bars for experimental data as well as for theoretical 
 calculations. 
 
\begin{figure}[htb]
\includegraphics[width=0.40\textwidth]{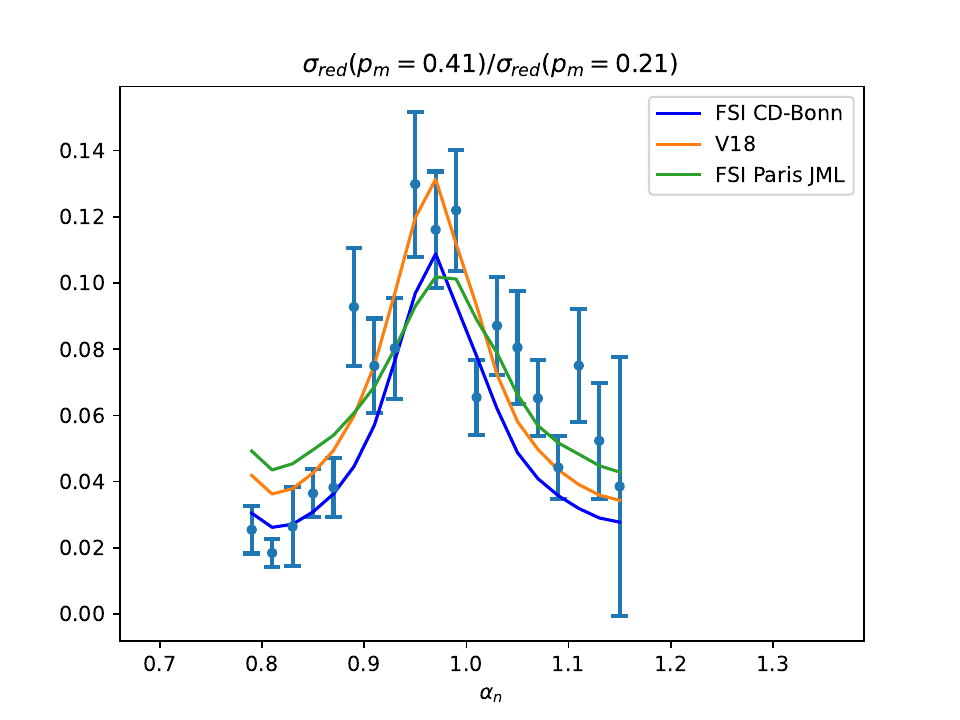}% 
\caption{The $\alpha_n$ dependence of  $R_\sigma$, defined in Eq.(\ref{Rsigma}).}
\label{fig:400_200_ratio}
\end{figure}

Finally,  in Fig.\ref{fig:400_200_ratio}  we present $\alpha_n$ dependence of the ratio of  reduced cross sections measured at double scattering ($p_m\approx 400$~MeV/c) and interference ($p_m\approx 200$~MeV/c) kinematics.
\begin{equation}
R_{\sigma} = {\sigma_{red}(p_m = 0.41)\over \sigma_{red}(p_m=0.21)}.
\label{Rsigma}
\end{equation}
The rationale of constructing such a ratio (see e.g. Ref.\cite{Li:2022uvf})  follows from the fact that it has maximal  sensitivity to the features of the $f_{pn}$ amplitude, since any 
modification of the latter will have opposite effects for cross sections 
measured at  $p_m\approx 400$~MeV/c and  $p_m\approx 200$~MeV/c.  For example if the
rescattering amplitude decreases  with $Q^2$ it will diminish the double scattering dominated cross section  that enters in the numerator  and increase the interference term dominated cross section that enters in the numerator of  ratio $R_\sigma$ (\ref{Rsigma}).

As Fig.\ref{fig:400_200_ratio} shows theoretical calculations describe reasonably well the first experimental evaluation of  
ratio $R_\sigma$.  It is interesting that as the figure shows the   above discussed intermediate $\Delta$-isobar contribution which is not included in the calculations largely cancels  out in the ratio.
This can be understood from the fact that the amplitude of  intermediate  
$\Delta$-isobar contribution is predominantly real which does not interfere with the PWIA term, thus having same effects in both numerator and denominator of ratio, $R_{\sigma}$.

\section{Conclusion and Outlook}
We presented the first analysis of the high $Q^2$ electrodisintegration of the deuteron in double scattering kinematics. 
Our main goal was to check   the conjecture that the double scattering subprocess can be unambiguously identified in the exclusive deuteron electrodisintegration reaction and its energy  (or $Q^2$) dependence reflects the energy dependence of $pn$ scattering  amplitude.
We took  advantage of the fact that the first moderately high $Q^2$ experiments are 
performed in the invariant energy range in which $pn$ scattering amplitude has noticeable energy dependence (Fig.\ref{fig:sigma_pn_tot}) ) thus providing sensitive test for the double scattering method of probing the
$pn$-amplitude.

The comparison  with theory shows a reasonable agreement with    the existing data confirming the sensitivity of double scattering processes to the energy dependence of the $pn$-scattering amplitude (Fig.\ref{fig:sdependence}).
This result indicates that the achieved accuracy in the description of the moderately high $Q^2$ data  in the nucleonic basis  allows us to extend  the considered reactions into a higher energy domain in which the onset of QCD degrees of freedom  is expected to modify rescattering processes due to  the dominance of small size color  neutral configurations in the wave function of  produced baryons at high $Q^2$.

Currently, the double scattering subprocess for the reaction (\ref{reaction}) is planned to be measured  for $Q^2$ as large  as 14~(GeV/c)$^2$\cite{HollyDoubleCT23} and it can significantly advance our understanding of the dynamics of the  $pn$ scattering amplitude in  the deuteron at large $Q^2$. 

 The deuteron disintegration can be extended also to deep inelastic kinematics for $d(e,e^\prime,N_r)X$  reactions in which  case the detection of  the spectator  nucleon 
 $N_r$ can be used to probe the reinteraction of the product of $\gamma^* N\to X$ scattering off the spectator nucleon\cite{Cosyn:2011jnm}. Such processes can be effective also for studying hadronization  dynamics  by isolating the rescattering  of  a hadron, h produced in $\gamma^* N\to h +X$ reaction off the spectator nucleon, $N_r$\cite{Cosyn:2017ekf}. The consideration of these processes in Electron-Ion Collider kinematics can significantly increase the effectiveness of these reactions allowing  the clear separation of current and target fragmentation regions.

\noindent {\bf Acknowledgment:}
This work is supported by United States Department of Energy grant under contract  DE-FG02-01ER41172 and DE-SC0013620.

\bibliography{references.bib}% Produces the bibliography via BibTeX.
\end{document}